\documentclass[12pt]{iopart}
\usepackage{iopams}

\newcommand{\vxi}{\vec{\xi}}
\newcommand{\vtau}{\vec{\tau}}

\newcommand{\vPi}{\vec{\Pi}}

\newcommand{\vomega}{\vec{\omega}}
\newcommand{\vP}{\vec{P}}
\newcommand{\vL}{\vec{L}}
\newcommand{\vn}{\vec{n}}
\newcommand{\vS}{\vec{S}}

\begin{document}
\title{Extended-body effects in cosmological spacetimes}
\author{Abraham I. Harte}
\address{Enrico Fermi Institute}
\address{University of Chicago, Chicago, IL 60637}
\ead{harte@uchicago.edu}

\begin{abstract}
We study the dynamics of extended test bodies in flat
Friedmann-Robertson-Walker spacetimes. It is shown that such objects
can usually alter their inertial mass, spin, and center-of-mass
trajectory purely through the use of internal deformations. Many of
these effects do not have Newtonian analogs, and exist despite the
presence of conserved momenta associated with the translational and
rotational symmetries of the background.
\end{abstract}

\date{June 14, 2007}

\vskip 1pc

\section{Introduction}

It is well-known that sufficiently small objects tend to fall along
the geodesics of whichever spacetime they inhabit. While it is
difficult to precisely state the limits of this result (see e.g.
\cite{Synge,GerochMotion, EhlersGeroch, PoissonRev, Damour, Nevin}),
some qualifications are clearly necessary. Even in Newtonian
gravity, the details of an extended test body's internal structure
can cause its center-of-mass trajectory to diverge from that of an
equivalent point particle. Inhomogeneities in the external field
effectively couple to the higher multipole moments of the mass
distribution to modify the center-of-mass motion.

In principle, this effect allows piloted spacecraft to partially
modify their trajectories simply by rearranging internal masses. A
particularly elegant example of this is a strategy whereby
artificial satellites can change their orbital parameters by
cleverly manipulating tethered masses \cite{Satellites}. Different
parts of the body effectively ``push'' or ``pull'' on local
gradients in the gravitational field. In the relativistic context,
test bodies interact with the background spacetime using their full
stress-energy tensors rather than just their mass distributions. The
control space available to alter trajectories using extended body
effects is therefore greatly enlarged. This has a number of
interesting consequences if very large stresses and internal momenta
can be maintained.

It allows, for example, a test body which starts at rest in a
homogeneous (but nonstationary) spacetime to accelerate purely
through the use of internal manipulations. No rocketry of any sort
is required. Even though there is no field anisotropy for the body
to push off of, it may still control its trajectory to some degree.
This is because a spacetime which may appear instantaneously uniform
to an observer comoving with the center-of-mass line needn't have
this property for the various frames associated with each of the
body's constituents. Sufficiently large internal velocities
therefore allow a temporal asymmetry in the geometry to have an
effect very similar to a spatial one. Related to this is an ability
for an initially nonrotating body to spin itself up by reducing its
center-of-mass velocity (appropriately defined).

There are other qualitatively new effects which occur in the
relativistic theory as well. Among these is the fact that a body's
rest mass is usually not conserved. This may be traced to changes in
what may be interpreted as the energy of the system. The
``gravitational potential energy'' can vary with a body's shape or
the nature of the applied field, for example. This affects its mass.

We show how to make these claims precise, and illustrate them using
the example of an uncharged extended test body inhabiting a
spatially-flat Friedmann-Robertson-Walker (FRW) spacetime. This
provides one of the first specific treatments of higher order
finite-size effects in the literature. Most previous work has
considered fully relativistic extended test bodies only as spinning
point particles \cite{SpinKerr,SpinUltra,SpinGravWave,SpinVaidya}.
These results assumed that quadrupole and higher order effects were
negligible, although this is not always physically reasonable. There
are many systems in which quadrupole effects dominate over those
associated with a particle's spin. While the general forms of these
corrections have been considered by several authors \cite{Mathisson,
Tulczyjew, Dix64, Taub, Madore, Dix70a, Dix74}, they do not appear
to have been previously studied in any specific systems.

\section{Laws of motion} \label{Sect:LawsMot}

The constructions used here derive from an extensive formalism
developed by Dixon \cite{Dix67, Dix70a, Dix74, Dix79} and others
\cite{EhlRud, CM, Isolated} to generalize useful concepts in
Newtonian gravitational mechanics to curved spacetimes. It considers
the behaviour of an extended body described by a nonsingular
stress-energy tensor $T^{ab}$ with spatially-bounded support $W$. To
briefly summarize, it has been shown that it is possible to convert
(without any approximation) the laws of motion
\begin{equation}
\nabla_{a} T^{ab} = 0 \label{divFree}
\end{equation}
into a handful of ordinary differential equations. These act on
objects which may be interpreted as the body's linear and angular
momenta. Simply put, a clear separation is made between those
components of $T^{ab}$ which are affected by stress-energy
conservation, and those which are purely constitutive. These latter
quantities may be identified as the body's higher multipole moments
(starting at the quadrupole). A number of other phenomenologically
interesting quantities naturally arise in this formalism as well.
These may be interpreted as a body's inertial mass, gravitational
potential energy, and so on.

For the present purposes, Dixon's formalism is most useful when
supplemented with a center-of-mass definition. Under mild
assumptions, it is possible to define a worldline $\Gamma$ which may
be shown to be uniquely defined, timelike, and within the convex
hull of $W$ \cite{CM}. It also reduces to the standard
center-of-mass definition in the appropriate limits. Rather than
solving for the full stress-energy tensor throughout the worldtube,
we only track $\Gamma$ and a few supplementary quantities necessary
to compute it.

Before proceeding with this, it is first necessary to define a
body's linear and angular momenta. Taking cues from the standard
flat-spacetime definitions of these quantities \cite{SRBook,MTW},
there should be some sense in which the momenta depend on time. At
any particular instant, it is reasonable to define them by
integrating the body's stress-energy tensor over an appropriate
spacelike hypersurface. To be more specific, suppose that the
center-of-mass worldline is parameterized by some function
$\gamma(s)$. $s$ then defines a natural notion of time. A spacelike
hypersurface $\Sigma(s)$ which contains $\gamma(s)$ may now be
associated with each point on $\Gamma$. The set of all such
hypersurfaces will be assumed to foliate the body's worldtube.

With these constructions in place, note that a particle's linear
momentum should be a vector field. In particular, let $p^{a}(s)$ be
a vector at $\gamma(s)$. Similarly, the angular momentum $S^{ab}(s)$
is taken to be a rank-2 skew-symmetric tensor at $\gamma(s)$. There
are an infinite number of reasonable definitions for momenta with
these properties. A unique choice may be made by supposing that any
Killing vector which a spacetime may possess implies that some
linear combination of $p^{a}$ and $S^{ab}$ is conserved. It is clear
that any Killing vector $\xi^{a}$ will imply the existence of a
conserved quantity
\begin{equation}
C_{(\xi)}  = \int_{\Sigma(s)}  \rmd \Sigma_{a} ( \xi_{b} T^{ab} ) ~.
\label{CDefine}
\end{equation}
We now suppose that $C_{(\xi)}$ may be constructed from the momenta
according to
\begin{equation}
C_{(\xi)} = p^{a}(s) \xi_{a}(\gamma(s)) + \frac{1}{2} S^{ab}(s)
\nabla_{a} \xi_{b}(\gamma(s)) ~. \label{CSum}
\end{equation}
This uniquely defines $p^{a}(s)$ and $S^{ab}(s)$ in terms of
$T^{ab}(x)$, $\gamma(s)$, and $\Sigma(s)$. The resulting expressions
are completely independent of $\xi^{a}$, so they may be adopted even
in the absence of any symmetries. Their detailed forms are not
needed here, and may be found in \cite{Dix70a, Dix79}.

Temporarily assuming that there again exists a Killing vector
$\xi^{a}$, (\ref{CSum}) indicates that some linear combination of
the force $F^{a}(s)$ and torque $N^{ab}(s)= N^{[ab]}(s)$ should
vanish. Differentiating (\ref{CSum}) with respect to $s$, one easily
finds that
\begin{equation}
F^{a} \xi_{a} + \frac{1}{2} N^{ab} \nabla_{a} \xi_{b} =0~,
\label{ForceVanish}
\end{equation}
where
\begin{eqnarray}
F^{a} := \delta p^{a}/\rmd s - \frac{1}{2} S^{bc} v^{d}
R_{bcd}{}^{a} \label{FDefine}
\\
N^{ab} := \delta S^{ab}/\rmd s - 2 p^{[a} v^{b]} ~, \label{NDefine}
\end{eqnarray}
and $v^{a}(s) = \dot{\gamma}^{a}(s) = \delta \gamma^{a}/ \rmd s$.
Note that the force automatically excludes the ``pole-dipole''
component of $\delta p^{a}/\rmd s$, while the definition of the
torque removes the term responsible for Thomas precession. Both
(\ref{FDefine}) and (\ref{NDefine}) make sense even in the absence
of any Killing vectors, so we consider these definitions to be
general.

Expressions for the force and torque may be derived from
(\ref{divFree}) in terms of $T^{ab}$ \cite{Dix74, Dix79}. If the
spacetime geometry inside each slice $W \cap \Sigma(s)$ is
sufficiently smooth (as discussed more precisely in \cite{Dix67}),
the resulting equations may be expanded as asymptotic series
involving successively higher multipole moments of the stress-energy
tensor. To lowest non-vanishing (i.e. quadrupole) order, it may be
shown that \cite{Dix70a, Dix74, Dix79}
\begin{eqnarray}
F_{a}(s) \simeq  - \frac{1}{6} J^{bcdf}(s) \nabla_{a} R_{bcdf}(
\gamma(s) ) \label{ForceQuad}
\\
N^{ab}(s) \simeq \frac{4}{3} J^{cdf [a}(s) R^{b]}{}_{fcd}( \gamma(s)
) ~. \label{TorqueQuad}
\end{eqnarray}
$J^{abcd} = J^{[ab][cd]} = J^{cdab}$ is the quadrupole moment of the
stress-energy tensor. It is defined precisely in \cite{Dix74},
although we simply note here that it does not satisfy any
differential equations as a consequence of (\ref{divFree}). All of
the higher multipole moments are constructed such that they share
this property. Their evolution equations are almost completely free,
and depend only on the type of matter under consideration.

%For now, we note that it is possible to construct $T^{ab}$ using $p^{a}$, $S^{ab}$ and an infinite set of other tensors defined along $Z$. These other tensors may be interpreted as the quadrupole and higher multipole moments. Remarkably, it is possible to adopt a definition of the higher moments such that (\ref{divT}) imposes differential constraints only on $p^{a}$ and $S^{ab}$; effectively the monopole and dipole moments. Clever definitions have therefore shown that the four partial differential equations of (\ref{divT}) are equivalent to ten ordinary differential equations which may be specified by giving explicit forms for $F^{a}$ and $N^{ab}$.

Up until now, we have been assuming that $\gamma(s)$ and its
associated hypersurfaces are known. These will now be chosen
uniquely using ``center-of-mass'' conditions. First let $n^{a}(s)$
be a unit timelike future-directed vector field defined on $\Gamma$.
In a sense, the hypersurfaces $\Sigma(s)$ are chosen to essentially
be hyperplanes orthogonal to $n^a(s)$. Their precise definition may
be found in \cite{Dix70a,Dix74,Dix79}. Regardless, our first
center-of-mass condition demands that the linear momentum be
proportional to this new vector field:
\begin{equation}
p^{a} = m n^{a} ~.
\label{CM1}
\end{equation}
$m(s)>0$ is some (not necessarily constant) proportionality factor
which we interpret as the object's inertial mass. This relation
suggests that $n^a$ be called the body's ``dynamical velocity.''
This is generally distinct from the ``kinematical velocity'' $v^a$.

Next, suppose that
\begin{equation}
p_{a} S^{ab} =0 ~.
\label{CM2}
\end{equation}
This reduces to the standard center-of-mass condition in flat
spacetime. It also leaves the angular momentum tensor with only
three independent components. Indeed, $S^{ab}$ may now be written in
terms of a single vector $S^{a}$ satisfying $n_{a} S^{a} =0$:
\begin{equation}
S^{ab} = \epsilon^{abcd} n_{c} S_{d} ~.
\label{SpinVect}
\end{equation}
Under mild assumptions, (\ref{CM1}) and (\ref{CM2}) uniquely
determine $n^{a}$ and $\Gamma$ throughout any given object
\cite{CM}. We call the worldline obtained with this method the
center-of-mass line.

Writing down equations of motion in the standard way now requires an
evolution equation for $\gamma(s)$. Despite the highly implicit
nature of the center-of-mass relations, it is possible to derive an
exact expression for $v^a - n^a$. Before describing this, it will
first be convenient to fix the scale of $s$ such that $v^a n_a =
-1$. In general, the time parameter used here therefore fails to be
the proper time of an observer moving on $\Gamma$. It is often very
close, however.

Defining the left and right duals of the Riemann tensor as
\begin{eqnarray}
R^{*}_{abcd} := \frac{1}{2} \epsilon^{pq}{}_{cd} R_{abpq}~,  \qquad
~{}^{*}\! R_{abcd} := \frac{1}{2} \epsilon_{ab}{}^{pq} R_{pqcd} ~,
\end{eqnarray}
it can now be shown that \cite{EhlRud}
\begin{equation}
\fl
\left( m^{2} + {}^{*}\! R^{*}_{bcdf} n^{b} S^{c} n^{d} S^{f}
\right) \left( n^{a} - v^{a} + \tau^{a} \right) = \epsilon^{abpq}
n_{p} S_{q} \left[ F_{b} - R^{*}_{bcdf} \left( n^{c} + \tau^{c}
\right) n^{d} S^{f} \right] ~. \label{CMEvolve}
\end{equation}
$\tau^{a}$ is that component of the torque which appears in the
expansion
\begin{equation}
N^{ab} = \epsilon^{abcd} n_{c} N_{d} + 2 m \tau^{[a} n^{b]} ~.
\label{TorqueExpand}
\end{equation}
This decomposition is unique if $n_a N^a= n_a \tau^a =0$.

The quadrupole moment $J^{abcd}$ can also be written in terms of
simpler tensor fields. Let \cite{EhlRud}
\begin{equation}
J^{abcd} =S^{abcd} - n^{[a} \pi^{b]cd} - n^{[c} \pi^{d]ab} - 3
n^{[a} Q^{b][c} n^{d]}  ~. \label{Quadrupole}
\end{equation}
$Q^{ab}(s) = Q^{(ab)}(s)$ may be interpreted as the quadrupole
moment of the mass distribution seen by an observer at $\gamma(s)$
with velocity $n^a(s)$. Similarly, $\pi^{abc} = \pi^{a[bc]}$ and
$S^{abcd} = S^{[ab][cd]}=S^{cdab}$ are essentially the body's
momentum and stress quadrupoles. These objects satisfy
\begin{equation}
n_{a} Q^{ab} = n_{b} \pi^{abc} = n_{a} S^{abcd} = 0 ~,
\end{equation}
and
\begin{equation}
\pi^{[a b c]} = 0 ~.
\end{equation}
%Given any second-rank tensor $\pi^{ab}$ satisfying $n_{a}\pi^{ab} =
%n_{b}\pi^{ab} = 0$, one may generate some $\pi^{abc}$ satisfying
%these constraints. In particular, let
%\begin{equation}
%\pi^{abc} = \left( \pi^{a}{}_{d} - \frac{1}{3} h^{a}{}_{d}
%\pi^{g}{}_{g} \right) n_{f} \epsilon^{dfbc} ~.
%\end{equation}
There are therefore 8 independent components of $\pi^{abc}$.
$Q^{ab}$ and $S^{abcd}$ each contain 6 components, leaving 20 for
the full quadrupole moment of the stress-energy tensor. This total
could also have been deduced immediately by noting that $J^{abcd}$
shares the same algebraic symmetries as the Riemann tensor.

The equations discussed in this section form the complete laws of
motion given by Dixon in the quadrupole approximation. They are not
the only such equations in the literature, however. Various
alternatives have been proposed \cite{Mathisson, Tulczyjew, Dix64,
Taub, Madore}, although each of these has drawbacks described in the
introduction of \cite{Dix74}. To summarize, some contain
calculational errors or inconsistent approximations. Others do not
include the ``full'' quadrupole moments of the stress-energy tensor.
Rather, some notions of the momentum and stress quadrupoles are
(explicitly or implicitly) neglected. Some also imply undesirable
results even in flat spacetime: e.g. changing masses or accelerating
center-of-mass worldlines. These issues were all resolved by Dixon's
formalism. For this reason and others, it will therefore be adopted
universally for the remainder of this article.

\section{Conservation laws}
\label{Conservation}

The previous section has developed enough of Dixon's formalism to
allow the study of test body motion in the quadrupole approximation.
The spatially-flat FRW spacetimes we consider admit at least six
linearly independent Killing vectors. Each of these implies the
existence of its own conserved quantity, which leaves four degrees
of freedom which are free to affect the evolution of $\Gamma$.
Suppose that we choose coordinates such that the metric takes the
form
\begin{equation}
\rmd s^{2} = - \rmd t^{2} + a^{2} (t) ( \rmd x^2 + \rmd y^2 + \rmd
z^2 ) ~. \label{metric}
\end{equation}
The scale factor $a(t)$ will be considered given. If Einstein's
equation is assumed to hold, it depends on the averaged density and
pressure of the background matter in the standard way.

Generators of the isometries associated with this FRW background may
be written down by inspection. First define the orthonormal tetrad
frame
\begin{equation}
  \mathbf{e}_{(0)} = \bpartial_t ~, \qquad \mathbf{e}_{(i)} = a^{-1}
  \bpartial_{i} ~, \label{Frame}
\end{equation}
where $i=1,2,3$. Using a 3-dimensional ``vector'' notation to denote
triads, the minimal set of Killing vector fields associated with
(\ref{metric}) may now be written in the form
\begin{equation}
\vxi = a \vec{e} ~, \qquad \vomega = \vec{R} \times \vxi ~.
\label{KillingVects}
\end{equation}
Here, $\vec{R}$ has been defined by the triplet $(x,y,z)$. The cross
product notation used to write down the rotational symmetries
$\vomega$ then has the standard interpretation. For example,
rotations about the $z$-axis are generated by the vector field
$\bomega_{(z)} = x \bpartial_{y} - y \bpartial_{x}$.

If (\ref{metric}) is assumed to hold throughout $W$, each of these
six Killing vectors implies the existence of a conserved quantity of
the form (\ref{CDefine}). Neglecting the body's self-field
essentially defines the test body approximation as it is used here,
although this assumption is not trivial. At first glance, it would
seem to imply that the cosmological fluid penetrates the body, and
that $T^{ab}$ is negligible compared to the stress-energy tensor of
the background. Most interesting systems do not satisfy these
constraints. Intuitively, though, they shouldn't have to. This
problem is a standard one, and we shall not attempt to address it
directly. It will instead be assumed that the methods used here can
be rigorously justified for a wide range of reasonable objects.

With this caveat out of the way, the constants associated with the
background symmetries -- usually referred to as the system's
conserved momenta -- may be directly related to Dixon's analogs of
these quantities. Slightly abusing the notation, the collection of
scalars associated with $\vxi$ through (\ref{CDefine}) will be
referred to as the body's global linear momentum $\vP =
(C_{(\xi_{(x)})}, C_{(\xi_{(y)})}, C_{(\xi_{(z)})})$. In contrast,
$p^a$ may be thought of as the system's \textit{local} linear
momentum. Directly evaluating (\ref{CSum}) using the translational
Killing vectors given in (\ref{KillingVects}) provides a direct
relation between these two objects. Simplifying it using
(\ref{CM1}), (\ref{SpinVect}), and the various normalization and
orthogonality conditions shows that
\begin{equation}
\vP = a \big( m \vn - H \vn \times \vS \big) ~. \label{PCalc}
\end{equation}
Here, $\vn$ and $\vS$ denote the triad components of the appropriate
vector fields with respect to $\vec{e}$. As is standard, the Hubble
parameter $H$ in (\ref{PCalc}) is defined by
\begin{equation}
H := \frac{ \rmd \ln a }{ \rmd t } ~. \label{HubDefine}
\end{equation}

The system's two notions of angular momenta may be understood in a
very similar way. Let those constants associated the rotational
Killing vectors $\vomega$ be denoted by $\vL$. As before, a direct
evaluation of (\ref{CSum}) immediately relates this object to $p^a$,
$S^a$, and $\gamma$. After simplifying with (\ref{PCalc}), it
reduces to
\begin{equation}
\vL - \vec{\gamma} \times \vP = \frac{ ( 1+ | \vn |^{2} ) \vS -  (
\vn \cdot \vS ) \vn }{ \left( 1 + | \vn |^{2} \right)^{1/2} } ~.
\label{LCalc}
\end{equation}
The dot product notation used here is just the standard Euclidean
definition. The norm $|\vn|^{2}$ is then $\vn \cdot \vn$. In
deriving (\ref{LCalc}), the Killing vectors must be evaluated at the
center-of-mass position $\gamma(s)$. We have therefore replaced
$\vec{R}$ by $\vec{\gamma} = (\gamma^{x}, \gamma^{y}, \gamma^z)$.
Notational conventions aside, the left-hand side of (\ref{LCalc})
may naturally be interpreted as the ``spin component'' of the global
angular momentum. In most cases, it differs very little from the
local spin angular momentum $\vS$. Note, however, that it does not
necessarily remain conserved. $\vP$ and $\vL$ are always fixed.
$\vec{\gamma}$ is not.

%Still it is important to note that it is only $\vL$ which
%necessarily remains conserved. As the center-of-mass moves,
%$\vL-\vec{\gamma} \times \vP$ may change.

\section{Forces and torques}

We now have explicit relations between the global and local notions
of momentum. Using the background tetrad, they may be used to write
$n^a$ and $S^a$ in terms of $\gamma$, $m$, $H$, $\vP$, and $\vL$.
This does not uniquely fix a body's center-of-mass position,
however. Its velocity $v^{a}$ satisfies (\ref{CMEvolve}), which also
requires knowledge of the force and torque. In the FRW spacetimes
considered here, $N^{ab}$ is entirely determined by the vector field
$\tau^{a}$ defined in (\ref{TorqueExpand}). With the exception of
the temporal component $F^{a} \nabla_{a}t$, this quantity also fixes
the force.

These statements follow from considering the consequences of the
conservation laws implied by the spacetime's symmetries. As
explained in Sec. \ref{Sect:LawsMot}, each of these implies that a
particular linear combination of the force and torque must vanish.
In this case, there are therefore six relations between the the
various components of $F^a$ and $N^{ab}$ which may be written down
without any detailed knowledge of the system. Directly applying
(\ref{ForceVanish}) for each of the Killing vectors in
(\ref{KillingVects}) shows that they have the form
\begin{equation}
\vec{F} = \frac{ m H \vtau }{ \sqrt{ 1+ |\vn|^{2} } }
\label{SpatialForce}
\end{equation}
and
\begin{equation}
\vec{N} = \frac{ m (\vn \times \vtau) }{ \sqrt{ 1+ |\vn|^{2} }} ~.
\label{SpatialTorque}
\end{equation}

There remain four free parameters here: $\vtau$ and $F^a \nabla_a
t$. These depend on the body's internal structure, and are
essentially free. There exists some extended body which may be
associated with almost any imaginable pair. The only restrictions
are imposed by energy conditions and the consistency of the test
body approximation.

In general, specific relations between a body's internal structure
and the force and torque acting on it may be found using the exact
integral expressions in \cite{Dix74, Dix79}. If the body under
consideration is much smaller than a Hubble length, its force and
torque may instead be approximated using multipole expansions.
Although the conservation laws used to derive (\ref{SpatialForce})
and (\ref{SpatialTorque}) are valid in the exact theory, they remain
exactly satisfied in any multipole approximation \cite{EhlRud}. It
is therefore fully consistent to consider the global momenta fixed
while using the quadrupole expressions for the force and torque
given in Sect. \ref{Sect:LawsMot}.

\section{Converting linear to rotational motion}

The two conservation laws (\ref{PCalc}) and (\ref{LCalc}) may be
used to write $\vn$ and $\vS$ in terms of $\vP$ and $\vL -
\vec{\gamma} \times \vP$. The first of these quantities is always
conserved, while the second is not. $\vec{\gamma} \times \vP$ may
vary as the body moves, which implies that the ``global spin'' may
change. This suggests that moving bodies are able to change their
local spin $S^a$ by appropriately manipulating their internal
structures.

Roughly speaking, the dynamics of the background spacetime mixes the
momenta. $\vP$ depends on both $\vS$ and $\vn$, for example. This
implies that if $\vn$ or $\vec{p} = m \vn$ can be changed, so will
$\vS$. Taking the norm of (\ref{PCalc}) illustrates this directly:
\begin{equation}
| \vec{p} | = \frac{ | \vP /a | }{ \sqrt{ 1 + |H \vS_{\perp}/m|^2} }
~. \label{pToP}
\end{equation}
Here, $\vS_{\perp}$ represents the component of $\vS$ orthogonal to
$\vn$. A larger spin may therefore be obtained by decreasing an
object's (local) linear momentum.

This equation only takes into account the conservation of linear
momentum, so one might expect (\ref{LCalc}) to rule out any such
effects. It does not. Suppose for simplicity that $\vP \cdot \vL
=0$, but $\vP \neq 0$. Then $\vn \cdot \vS =0$ regardless of the
particle's internal dynamics. It follows that
\begin{equation}
\vS = \vS_{\perp} = \frac{ \vL - \vec{\gamma} \times \vP }{ \sqrt{ 1
+ | \vn |^2 }} ~. \label{lToS}
\end{equation}
In this case, the magnitude of the angular momentum vector coincides
with that of the triad components displayed here: $S^a S_a = \vS
\cdot \vS$. This clarifies the interpretation of (\ref{pToP}).

Combining (\ref{pToP}) with (\ref{lToS}) allows either $|\vS|$ or
$|\vn|$ to be expressed in terms of $m$, $H$, $|\vP/a|$, and $|\vL -
\vec{\gamma} \times \vP|$. Although straightforward to derive, the
resulting expressions are quite lengthy. We therefore note only that
to lowest order in $\vP$, they show that
\begin{equation}
| \vS_{\perp} | = | \vL - \vec{\gamma} \times \vP | \left[ 1  +
o(|\vP/ma|^2) \right] ~. \label{SExpand}
\end{equation}
This can also be deduced directly by noting that $|\vn| \simeq
|\vP/ma|$ in this limit. The denominator of (\ref{lToS}) may
therefore be expanded directly in the global momentum.
(\ref{SExpand}) follows trivially.

%For example, deriving $|\vn| = |\vec{p}|/m$ requires substituting the magnitude of (\ref{lToS}) in
%to (\ref{pToP}).

Regardless, this illustrates that the magnitude of a given
particle's spin depends on its center-of-mass position. In the
simplest cases, $\rmd \vec{\gamma}/\rmd s$ will be nearly
proportional to $\vP$, implying that $|\vec{S}_{\perp}|$ remains
constant. More generally, though, extended-body effects allow
significant variations in the spin.

It is interesting to ask whether these changes are generic. Consider
a particle which does not rotate, so $\vS =0$. In this case,
(\ref{PCalc}) and (\ref{LCalc}) imply that
\begin{equation}
\vec{\gamma} \times \vP = \vL ~, \qquad \vn = \vP/ma ~.
\end{equation}
Differentiating the first of these expressions with respect to $s$
shows that $\vec{v} \times \vn = 0$. This condition is both
necessary and sufficient to ensure that the local angular momentum
vanishes. $\vec{v}$ must therefore remain proportional to $\vn$.
Using (\ref{CMEvolve}) finally shows that $\vtau \propto \vP$. As
might have been expected directly from (\ref{SpatialTorque}), a
nonspinning particle can only remain in that state if its internal
structure is arranged such that the temporal torque remains
proportional to the conserved momentum. Other possibilities alter
the body's direction of travel, and inevitably impart a spin in the
process.

This statement assumes that $\vS$ initially vanishes. Cases where
$\vtau \propto \vn$ always have special significance, however (at
least when $\vP \cdot \vL =0$). Differentiating (\ref{lToS})
directly, it is clear that $\rmd \vS_{\perp} / \rmd s$ involves
terms proportional to both $\vS_{\perp}$ and $\vec{v} \times \vP$.
The second of these is more interesting.

The center-of-mass velocity may be found using (\ref{CMEvolve}) and
(\ref{SpatialForce}). Here, choosing $\tau \propto \vn$ implies that
\begin{equation}
n^{a}-v^a+\tau^a \propto - \epsilon^{abpq}n_{p} S_{q} R^{*}_{bcdf}
n^c n^d S^f ~.
\end{equation}
It may then be shown that $\vec{v} \propto \vn$. Computing the spin
therefore requires knowledge of $\vn \times \vP$. This is most
easily obtained by using (\ref{PCalc}) and (\ref{LCalc}) to show
that
\begin{equation}
a m \vn = \frac{ \vP + (H/m) \vP \times \vS_{\perp} }{ 1 + |H
\vS_{\perp}/m|^2 } ~.
\end{equation}
It follows that whenever $\vtau \propto \vn$,
\begin{equation}
\rmd \vS_{\perp}/\rmd s  \propto \vS_{\perp} ~.
\end{equation}
The direction (but not necessarily the magnitude) of the local
angular momentum therefore remains fixed in all such cases.

\section{Zero-momentum particles}

The detailed behaviour of extended bodies with arbitrary momenta can
be quite complicated. Suppose for simplicity that
\begin{equation}
  \vP =0 ~. \label{zeroP}
\end{equation}
This is the unique condition which locks a particle's motion to that
of the background fluid in the monopole and dipole approximations.
Interesting effects appear at the quadrupole level, and are
surprisingly rich even in this simple case.

The spin behaviour discussed in the previous section is not
retained, however. Dotting (\ref{PCalc}) with $\vn$ implies that
$\vec{p}$ must vanish whenever $\vP$ does. It then follows from
(\ref{LCalc}) that $\vS =\vL$ in this case. The local and global
angular momenta agree exactly, so the local spin is conserved. The
choice (\ref{zeroP}) therefore isolates those extended body effects
which are as independent of the angular momentum as possible.

Contracting (\ref{CMEvolve}) with $e_{(0)}^a$ and recalling that
$n_{a} \tau^{a} =0$ now shows that up to a possible additive
constant, the worldline parameter $s$ is simply the cosmological
comoving time $t$. The remainder of (\ref{CMEvolve}) then yields the
equation-of-motion
\begin{equation}
m \frac{ \rmd \vec{\gamma} }{\rmd t} = \frac{ m \vtau + H \left[ \vL
\times \vtau + (H/m) (\vL \cdot \vtau ) \vL \right] }{ a \big( 1 + |
H \vL/m |^{2} \big) } ~. \label{CMandTau}
\end{equation}
This expression reduces to $\vec{v} = a  \rmd \vec{\gamma}/ \rmd t =
\vtau$ when $\vL \propto \vtau$. In other words, the coordinate
velocity matches the coordinate (not frame) components of $\tau^a$.
Unless the angular momentum is extremely large, this will remain an
excellent approximation for the center-of-mass velocity of any
system satisfying $\vP=0$.

The temporal torque $\vtau$ -- and therefore the velocity $\vec{v}$
-- may be controlled by a body allowed to manipulate its internal
structure. Combining (\ref{TorqueExpand}), (\ref{Quadrupole}), and
(\ref{TorqueQuad}), it can be shown that in the quadrupole
approximation,
\begin{equation}
m \vtau \simeq \frac{2}{3} \dot{H} \vPi ~. \label{TauQuad}
\end{equation}
Here, $\dot{H} := \rmd{H}/\rmd t = \rmd{H}/\rmd s$ and
\begin{equation}
\Pi^{a} :=\pi^{ba}{}_{b} ~.
\end{equation}
This trace of the momentum quadrupole may point in any direction
whatsoever (orthogonal to $n^a$), so an object with vanishing global
momentum could have complete control over its direction of motion.
The magnitude of this quantity limits the magnitude of the
center-of-mass velocity. Indeed,
\begin{equation}
| \vec{v} | \simeq |\vtau| \left( \frac{ 1 + (H/m)^{2} ( \vL \cdot
\vtau)^{2}/ | \vtau |^2 }{ 1 + | H \vL /m |^{2} } \right)^{1/2} ~.
\label{CMVelocity}
\end{equation}
It follows that $|\vec{v}| \lesssim |\vtau|$ in the quadrupole
approximation. These two quantities coincide only when $\vS \times
\vPi=0$ or $\vS \cdot \vPi=0$. In other cases, a nonzero spin will
reduce the magnitude of the particle's ``drift'' velocity.

$\vtau$ depends on the mass parameter $m$, so $\vec{v}$ does as
well. In general, (\ref{FDefine}) and (\ref{CM1}) may be used to
show that
\begin{equation}
\frac{\rmd m}{\rmd s} = R^{*}_{abcd} n^{a} v^{b} n^{c} S^{d} -n_{a}
F^{a} ~, \label{MassChange}
\end{equation}
Adapted to the present case, this reduces to
\begin{equation}
\frac{ \rmd m }{ \rmd t} \simeq - \frac{1}{2} \frac{ \rmd }{\rmd t}
( \dot{H} + H^{2} ) Q^{a}{}_{a} + \frac{1}{3} \frac{ \rm{d}}{ \rm{d}
t} ( H^{2} ) S^{ab}{}_{ab} ~. \label{DmDt}
\end{equation}
Mass changes therefore depend only on the quadrupole moments of the
body's mass and stress distributions. They are independent of the
momentum quadrupole. This quantity only affects the particle's
velocity with respect to the cosmological rest frame, which is
itself independent of $Q^{ab}$ and $S^{abcd}$. Each of these three
components of the full quadrupole moment may be varied
independently. By construction, stress-energy conservation does not
require any coupling between them. In principle, a
properly-engineered spacecraft could therefore control its mass and
velocity separately.

The coefficients in front of the moments in (\ref{DmDt}) are written
so as to make this equation trivial to integrate if the traces
$Q^{a}{}_{a}$ and $S^{ab}{}_{ab}$ remain constant. That this is
possible is not an accident. In the exact theory, the mass may be
written as a sum of terms which are naturally interpreted as the
total internal energy $m_{\rm{int}}$, the gravitational potential
energy $\Phi$, and the rotational kinetic energy $E_{\rm{rot}}$ of
the body \cite{Dix79}:
\begin{equation}
m = m_{\rm{int}} + \Phi + E_{\rm{rot}} ~. \label{MassExpand}
\end{equation}

The rotational contribution here may be defined in terms of $S^a$
and an angular velocity derived from an inertia tensor based on
$Q^{ab}$ \cite{Dix70a, Dix79}. If the quadrupole moment remains
fixed in an appropriate sense, the fact that $S_{a} S^{a} = | \vL
|^{2}$ cannot change ensures that $E_{\rm{rot}}$ remains constant.

More generally, it is probably most straightforward to study the
behaviour of the sum $m_{\rm{int}} + E_{\rm{rot}} = m-\Phi$. This
may be found by noting that in the quadrupole approximation, the
gravitational potential energy reduces to \cite{Dix70a}
\begin{equation}
\Phi \simeq \Phi_{0} + \frac{1}{6} J^{abcd} R_{abcd} ~,
\end{equation}
where $\Phi_{0}$ is an arbitrary constant. Applying this to the
present case, one sees that
\begin{equation}
\Phi \simeq \Phi_{0} - \frac{1}{2} ( \dot{H} + H^{2} ) Q^{a}{}_{a} +
\frac{1}{3} H^{2} S^{ab}{}_{ab} ~.
\end{equation}
Comparison with (\ref{DmDt}) now shows that the internal and
rotational energies must evolve according to
\begin{equation}
\frac{ \rmd (m - \Phi) }{ \rmd t} = \frac{1}{2} ( \dot{H} + H^{2} )
\frac{ \rmd Q^{a}{}_{a}}{ \rmd t} - \frac{1}{3} H^{2} \frac{\rmd
S^{ab}{}_{ab}}{\rmd t} ~.
\end{equation}
This effectively describes the rate at which energy is inductively
absorbed from the gravitational field.

It has been proven in the exact theory that $m_{\rm{int}}$ remains
constant for any ``rigid'' body; i.e. one whose multipole moments do
not change with respect to an appropriate co-rotating tetrad
\cite{Dix79}. In this special case, full rigidity is not required
even of the quadrupole moment. The internal energy is conserved for
any object where $Q^{a}{}_{a}$ and $S^{ab}{}_{ab}$ remain constant.
This condition is weaker than one requiring that all tetrad
components of $J^{abcd}$ remain constant.

When evaluating (\ref{CMVelocity}) and (\ref{DmDt}) in specific
spacetimes, it is useful write $\dot{H}$ and $\ddot{H}$ in terms of
quantities with more direct physical interpretations. For the metric
(\ref{metric}), Einstein's equation is equivalent to \cite{MTW}
\begin{eqnarray}
3 H^{2} &= 8 \pi \rho + \Lambda \label{HFluid}
\\
3 \ddot{a}/a &= - 4 \pi (\rho + 3 p) + \Lambda ~.
\end{eqnarray}
$\rho$ and $p$ are respectively the density and pressure of the
background matter, while $\Lambda$ represents the cosmological
constant. Suppose for simplicity that the cosmological fluid obeys
an equation of state of the form
\begin{equation}
p = w \rho ~.
\end{equation}
Also introduce the density parameter
\begin{equation}
\Omega := 8 \pi \rho/3 H^{2} =  1 - \Lambda/3 H^{2} ~.
\end{equation}

Ignoring the particle's spin, (\ref{CMVelocity}) and (\ref{DmDt})
now reduce to
\begin{equation}
m \vec{v} \simeq - (1+w) \Omega H^{2} \vPi ~, \label{CMVelocity2}
\end{equation}
and
\begin{equation}
\dot{m} \simeq - (1+w) \Omega H^{3} \left[ \frac{3}{4} (1+3 w)
Q^{a}{}_{a} + S^{ab}{}_{ab} \right] ~. \label{DmDt2}
\end{equation}
If $\Omega$ vanishes or $w=-1$, the spacetime is de Sitter. In these
cases, both an object's mass and position remain fixed for all time
regardless of its internal structure. This could have been deduced
directly using the fact that de Sitter spacetimes possess an extra
four Killing vectors beyond the six in (\ref{KillingVects}).
Together, these ensure that all components of the force and torque
must vanish. There exist ten conserved quantities, among which is
$m$. The center-of-mass line is always a geodesic in this case, and
the angular momentum will be parallel-propagated along that
geodesic.

Generically, though, the object's mass will almost always change.
Nearly any imaginable object has a nonzero mass quadrupole. Standard
energy conditions should then imply that $Q^{a}{}_{a} \neq 0$.
Excluding the de Sitter case, the first term in (\ref{DmDt2}) can
only vanish if $w= - 1/3$. This possibility only marginally
satisfies the strong energy condition, so it is unlikely to be
physically relevant.

Still, these effects are exceedingly small even in the most
favorably-chosen systems. If the characteristic diameter of $W$ in
the center-of-mass frame is of order $D(s)$, the maximum $|\vec{v}|$
which the particle may attain by the method we've described is of
order $(D/10^{10}~\rm{yr})^{2}$ in the present universe. This value
can only become significant for objects on the scale of galactic
superclusters. It is a best-case estimate, however. The relatively
low internal velocities of most types of astrophysical matter makes
it very unlikely that there exists any observable drift velocity
even in these systems.

The situation improves slightly if we no longer assume that $\vP=0$.
For small momenta, $\vec{v}$ can be shown to depend on the mass
quadrupole as well as $\vPi$. This eliminates the need for a
complicated internal velocity distribution. All that's required is
an appropriate mass distribution together with some bulk motion. In
either case, though, the maximum change in the center-of-mass
velocity still scales as $(\textrm{characteristic speed}) \times
(D/10^{10}~\rm{yr})^2$. The ultra-relativistic limit $|\vP|/ma \gg
1$ is more interesting. If the magnitude of $J^{abcd}$ is kept fixed
with respect to a center-of-mass frame, boosting the quadrupole
moment appropriately leads to a considerable amplification of the
effects we've discussed. Of course, this case isn't particularly
relevant to astrophysical systems either.

Mass shifts are potentially more interesting. Although the
fractional \textit{rate} of mass change will be very small, no
special structure is required to maintain it (and its sign). For
very large objects, the total change in $m$ over a significant
fraction of the universe's history could be significant. Suppose
that $Q^{a}{}_{a}$ remains constant, and ignore the stress
quadrupole. (\ref{DmDt}) and (\ref{HFluid}) then imply that in a
matter-dominated universe, an object's mass will change by an amount
\begin{equation}
m_{\mathrm{f}} - m_{\mathrm{i}} \simeq - \frac{1}{4} Q^{a}{}_{a}
H_{\mathrm{f}}^{2} \Big[ \big( a_{\mathrm{f}} / a_{\mathrm{i}}
\big)^{3} - 1 \Big] ~. \label{massChange}
\end{equation}
Given that this scales as the cube of the redshift factor between
the initial and final states, relatively large changes may exist
even though $Q^{a}{}_{a} H_{\mathrm{f}}^{2} \ll m$. For example, a
galactic supercluster ($D \sim 10^{8}~\rm{yr}$ today) formed at a
redshift of $\sim 10$ might have lost as much as $5\%$ of its mass
by the present time. Of course, this model is very crude. Among
other problems, there is no reason to expect that $Q^{a}{}_{a}$
would remain even approximately constant for so long. It is also not
clear how the notion of mass used here -- which is essentially
inertial -- relates to the gravitational masses so fundamental to
astronomical observations. At least in stationary spacetimes, these
two types of mass are not identical \cite{Isolated}. Understanding
the analogous relations in the present case would require further
investigation.

Mass loss effects have been noted before in cosmological spacetimes
for freely-falling point particles endowed with a scalar charge
\cite{HarteScalar, PoissonScalar}. In that case, the effect is due
to an object's own scalar radiation reflecting back towards itself.
This arises from the failure of Huygen's principle in curved
spacetimes. Such exotic mechanisms are apparently not required to
induce mass changes in freely-falling particles (even rigid ones).

\section{Discussion}

Despite the simplicity of the systems studied here, a number of
interesting extended-body effects were found to exist. Even in the
presence of linear and angular momentum conservation laws, it was
shown that bodies can control the magnitudes of their center-of-mass
acceleration and spin using purely internal processes. This
illustrates some of the richness of finite-size effects in general
relativity. Similar phenomena surely exist in other spacetimes, and
their magnitudes are likely to be much larger for objects with
realistic dimensions.

It's also worthwhile to note that the concept of controllable motion
emphasized here doesn't occur in simpler approximation schemes. If
the quadrupole and higher multipole moments of a body are ignored in
its laws of motion, one recovers the Papapetrou equations
(occasionally called the Mathisson-Papapetrou or
Mathisson-Papapetrou-Dixon equations). These follow from
(\ref{divFree}), and involve no free parameters once a definition is
fixed for the center-of-mass. The motion of a spinning test body in
a given background is completely determined by its initial
conditions.

This no longer remains true once the effects of higher multipole
moments are taken into account. Dixon has defined these moments in
such a way that stress-energy conservation does not affect their
evolution. Their time-dependence is made unique by specifying
constitutive relations which depend on the type of material under
consideration. These are essentially equations of state, and are
simplest to consider in the case of an ideal programmed mechanical
device. The internal structure may then be considered a given
function of time.

Applying the equations of motion we've derived to more standard
``passive'' materials can be relatively complicated. This requires
specifying an evolution equation for the quadrupole moment in terms
of the various other bulk quantities relevant to the motion. For
most realistic materials, all of the multipole moments couple to
each other. It may not always be possible to ignore these couplings,
in which case the body is probably best modeled using the full
partial differential equations of continuum mechanics.

\ack

I would like to thank Pablo Laguna for useful comments. This work
was also largely inspired by discussions with Deepak Vaid.

%In the quadrupole approximation, the spatial frame components of
%$\tau^a$ with respect to (\ref{BaseFrame}) are (without using any
%slow-motion approximations)
%\begin{equation}
%\tau^{(i)} = \frac{4}{3} ( 1 + | \vn |^2 ) \dot{H}
%J^{(i)(a)}{}_{(0)(a)} /m ~.
%\end{equation}
%The quadrupole moment here is evaluated in a frame adapted to the
%background. It is much more natural to express it in terms of one
%comoving with the body. Define this intrinsic frame by an
%orthonormal tetrad with temporal component
%$\bar{e}^{a}_{(\bar{0})}=n^a$. Denote the transformations from one
%frame to another by
%\begin{equation}
%\Gamma^{(a)}{}_{(\bar{a})} := e^{(a)}_{b} \bar{e}^{b}_{(\bar{a})} ~.
%\end{equation}
%Then
%\begin{equation}
%\tau^{(i)} = \frac{4}{3} (1 + |\vn|^2 ) \dot{H}
%\Gamma^{(i)}{}_{(\bar{a})} \Gamma_{(0)}{}^{(\bar{c})} J^{(\bar{a})
%(\bar{b})}{}_{ (\bar{c}) (\bar{b}) }
%\end{equation}

%Suppose for simplicity that only the quadrupole moment of the mass
%distribution is significant. Using (\ref{Quadrupole}), it is then
%possible to show that
%\begin{equation}
%\tau^{(i)} = - (1 + |\vn|^2) (\dot{H} / m) \left( \Gamma^{(i)}{}_{
%(\bar{i})} \Gamma_{(0)}{}^{ (\bar{j}) } Q^{ (\bar{i}) }{}_{
%(\bar{j})} + \Gamma^{(i)}{}_{ (\bar{0}) } \Gamma_{(0)}{}^{ (\bar{0})
%} Q^{ (\bar{i}) }{}_{ (\bar{i})} \right) ~.
%\end{equation}
%At low speeds,
%\begin{equation}
%\tau^{(i)} \simeq (\dot{H}/m) \left( n^{(j)}
%\delta^{(i)}_{(\bar{i})} \delta^{ (\bar{j}) }_{ (j)} Q^{ (\bar{i})
%}{}_{ (\bar{j}) } - n^{(i)} Q^{ (\bar{j}) }{}_{ (\bar{j}) } \right)
%~.
%\end{equation}

\end{document}